# Layered Based Augmented Complex Kalman Filter for Fast Forecasting-Aided State Estimation of Distribution Networks

Mehdi Shafiei, *Student Member, IEEE*, Gerard Ledwich, *Senior Member, IEEE,* Ghavameddin Nourbakhsh, *Member, IEEE*, Ali Arefi, *Senior Member, IEEE*, Houman Pezeshki, *Member, IEEE*

*Abstract*— In the presence of renewable resources, distribution networks have become extremely complex to monitor, operate and control. In addition, with the high cost of measurement devices, communication infrastructure and data handling, there is a strong motivation to design and employ suitable distribution state estimation (DSE) algorithms. Furthermore, for the real time applications, active distribution networks require fast real-time DSE to design control and protection algorithms. Forecasting-aided state estimator (FASE), deploys measured data in consecutive time samples to refine the state estimate. Although most of the DSE algorithms deal with real and imaginary parts of distribution networks' states independently, we propose a new non-iterative complex DSE algorithm based on augmented complex Kalman filter (ACKF) which considers the states as complex values. In the proposed method, the combination of measured and pseudo data is deployed in the measurement vector to make distribution network observable. In case of real-time DSE and in presence of a large number of customer loads in the system, employing DSEs in one single estimation layer is not computationally efficient. Consequently, our proposed method performs in several estimation layers hierarchically as a Multi-layer DSE based on ACKF (DSE-MACKF). In the proposed method, a distribution network can be divided into one main area and several subareas. The aggregated loads in each subarea act like a big customer load in the main area. Load aggregation results in a lower variability and higher cross-correlation. This increases the accuracy of the estimated states in the main estimation layer. Additionally, the proposed method is formulated to include unbalanced loads in low voltage (LV) distribution networks. The results are evaluated and examined on two real distribution networks. The effectiveness of the proposed method is discussed using several criteria such as computational time, standard deviation, and maximum and average voltage error.

## I. Introduction

STATE estimation (SE) has been used extensively in transmission network operation, control and planning studies [1]. In recent years, the emerging distributed energy resources (DERs) have created demand for SE in distribution networks. Distribution state estimation (DSE) is used in the distribution networks for various applications such as Volt/Var control, feeder reconfigurations, short term planning, demand response management and power market operation [2-4]. The most common DSE is designed based on the weighted least square (WLS) method, as detailed in [5]. The main limitation of WLS-based works is requiring extensive real measurements with fast communication platforms, which is not cost-effective in distribution networks.

Traditionally, SE algorithms are formulated with a static nature in which states are estimated based on single instantaneous measurements, and previous measured values are incorporated in the estimation. However, due to high cross-correlations, valuable information can be extracted from previous measured data to be used in forecasting-aided SE (FASE) methods. In general, FASE algorithms are used to detect unexpected variation in the system states in control and protection analysis, network configuration errors and bad data detection [6]. For instance, in the security and control analysis, a recursive FASE method based on measurements is reported in [7], and also a comprehensive study of this method is detailed in [6, 8]. WLS as a snapshot FASE algorithm with high resolution measured data is applied in [9, 10], but it requires a high number of measurement devices which makes it a costly solution. In [11], it is shown that Kalman filter as the most common time series forecasting-aided state estimator [12] has a better performance compared with WLS for distribution networks. In [13], two decoupled FASE algorithms are proposed for estimating voltage magnitude and voltage angle independently. These algorithms are not efficient due to their high computational cost and ignoring the dependencies between magnitude and angle estimation noises. A complex formulation for the state estimator can potentially improve the accuracy of estimation [14].

As noted before, operating a fully measured distribution network is not cost-effective, however, the lack of real measurements in distribution networks makes DSE under-determined, unless pseudo values are used for unmeasured points [15]. Although using pseudo data can make the system fully observable for state estimation, it introduces significant errors which decreases the accuracy of the estimated states [16]. Several methods have been devised to increase the accuracy of the pseudo measured data, such as a Gaussian model to represent the error associated with pseudo measured data in [17]. Although these methods increase the accuracy of pseudo measured data, still require a very large number of measurements with advanced communication platforms. These requirements not only are expensive, but also impose a high computation time, while state estimators used for designing control and protection algorithms need to capture the dynamic of the system in a very short period of time [18].

Distribution networks contain an extremely large number of customer nodes, and it is not computationally efficient to process state estimation in a single layer. Instead, the network can be divided into several subareas, where DSE is carried out in sequence or parallel [19]. In a multi-layers state estimation approach, several factors can be considered to determine the boundaries of the subareas such as overlapping buses, coordination and synchronization [20]. A multi-layers DSE is presented in [16, 17] considering parallel and series zones. In parallel zones, DSE is employed for several zones simultaneously allowing for a lower computation time. Similarly, in series zones, network schematic matrix reduction leads to a higher computational efficiency. Although this multi-layers state estimation method reduces the computational time, it requires a significant number of real measurements, making them economically unattractive and computationally inefficient for real distribution networks.

In order to address the low accuracy and computational efficiency of the existing forecasting-aided DSE models, inspired by [14], this paper proposes a new single iteration Multi-layer model for DSE based on Augmented Complex Kalman Filter (ACKF) (DSE-MACKF). We show that the continuous variations in customer loads with a short sampling time make variations like white noise. This allows us to consider the injected current as the estimator states in the proposed DSE-MACKF. In distribution networks with a very low number of measurement devices, our proposed method substantially decreases the estimation error. To decrease the simulation time in distribution networks with a massive number of customer loads, several estimation layers are considered to divide the network into one main area and several subareas. We calculate the contribution of each subarea to the measured current on the secondary of the MV/LV transformer based on historical data and in real-time compute the injected currents of subareas. To get a more accurate estimation of subareas' contributions, the customer loads phase adjacent aggregation is considered to reduce variability and increase the spatial correlation within subareas. The proposed approach as a real-time state estimator can be applied to both balanced and unbalanced distribution networks with complex network states.

In summary, the main contributions of this paper are:
1. Formulating a new complex non-iterative FASE algorithm for both balanced and unbalanced distribution networks with a very low number of measurement devices. This paper shows that how the use of the time series measurements can substantially reduce the error of estimation by continually refining the estimated states based on available data.
2. Phase adjacent aggregation is employed to divide the network into several estimation layers, where the pseudo-injected current of each layer is updated by pseudo scaling factors.

The efficiency of the proposed approach is evaluated based on two real LV distribution networks with 15% to 25% PV penetration. Simulation results in Section V confirm that the proposed method considerably decreases state estimation error and computational time.

## II. FORECASTING-AIDED COMPLEX STATE ESTIMATOR: BASICS AND FORMULATION

The complex nature of the bus voltages, branch and injected currents as the system states requires a complex formulation of DSE algorithms for an efficient characterization. Although, estimating the real and imaginary parts of the states independently decreases the complexity of the algorithm, it ignores the interaction between real and imaginary parts. In the following subsections, first the general formulation of ACKF is given. Then, we build our proposed DSE-ACKF framework on the basis of direct load flow formulation.

### A. Augmented Complex Kalman Filter

A general linear state-space model is given as [21]:

$$\begin{aligned} x_i &= F_{i-1} x_{i-1} + w_i \\ y_i &= H_i x_i + n_i \end{aligned} \quad (1)$$

where $x_i$ and $y_i$ are the complex state and measurement vectors of the system at discrete time $i$, respectively. $F_{i-1}$ and $H_i$ denote state transition and observation matrices, and $(w_i)$ and $(n_i)$ represent white noises.

In a strictly linear distribution network model with complex state and measurement vectors, (1) can be rewritten in a so-called augmented complex vectors format as [14]:

$$\begin{aligned} x_i^a &= F_{i-1}^a x_{i-1}^a + w_i^a \\ y_i^a &= H_i^a x_i^a + n_i^a \end{aligned} \quad (2)$$

where $(.)^a$ represents the augmented complex vectors; $x_i^a = \begin{bmatrix} x_i \\ x_i^* \end{bmatrix}$, $y_i^a = \begin{bmatrix} y_i \\ y_i^* \end{bmatrix}$, $F_i^a = \begin{bmatrix} F_i & 0 \\ 0 & F_i^* \end{bmatrix}$, $H_i^a = \begin{bmatrix} H_i & 0 \\ 0 & H_i^* \end{bmatrix}$, $w_i^a = \begin{bmatrix} w_i \\ w_i^* \end{bmatrix}$ and $n_i^a = \begin{bmatrix} n_i \\ n_i^* \end{bmatrix}$, with $(.)^*$ as the complex conjugate operation.

In the state-space model, it is assumed that the state and measurement noises are uncorrelated with zero mean [22]. The augmented state and measurement noise covariance matrices $Q_i^a$ and $R_i^a$ can be calculated using (3):

$$\begin{aligned} Q_i^a &= E(w_i^a w_i^{aH}) = \begin{bmatrix} \Gamma_Q & C_Q \\ C_Q^* & \Gamma_Q^* \end{bmatrix} \\ R_i^a &= E(v_i^a v_i^{aH}) = \begin{bmatrix} \Gamma_R & C_R \\ C_R^* & \Gamma_R^* \end{bmatrix} \end{aligned} \quad (3)$$

where $E(.)$ and $(.)^H$ are the mean and transpose-complex conjugate operators, respectively. $\Gamma_Q$ and $\Gamma_R$ are the covariance matrices, and $C_R$ and $C_Q$ are pseudocovariance matrices for complex state and measurement noises.

After modelling a linear system in a complex state-space format, and introducing state and measurement noises, ACKF can be formulated through the following three steps:

1) *State and covariance matrix initialization:*

$$\begin{aligned} \hat{x}_{0|0} &= E(x_0) \\ P_{0|0}^a &= E\{(x_0^a - E(x_0^a))(x_0^a - E(x_0^a))^H\} \end{aligned} \quad (4)$$

2) *Updating states and covariance matrix:*

$$\begin{aligned} \hat{x}_{i|i-1} &= F_{i-1} \hat{x}_{i-1|i-1} \\ P_{i|i-1}^a &= F_{i-1}^a P_{i-1|i-1}^a F_{i-1}^{aH} + Q_i^a \end{aligned} \quad (5)$$

3) *Updating states based on measurements:*

$$\begin{aligned} G_i^a &= P_{i|i-1}^a H_i^{aH} (H_i^a P_{i|i-1}^a H_i^{aH} + R_i^a)^{-1} \\ \hat{x}_{i|i} &= \hat{x}_{i|i-1} + G_{i_{11}}(y_i - H_i \hat{x}_{i|i-1})k + G_{i_{12}}(y_i^* - H_i^* \hat{x}_{i|i-1}^*) \\ P_{i|i}^a &= (I - G_i^a H_i^a) P_{i|i-1}^a \end{aligned} \quad (6)$$

where $\hat{x}_{i|i}$ shows a posteriori states at time $i$ giving observation including at time $i$, $P^a$ denotes the augmented state covariance matrix, and $G^a$ is the Kalman gain. $G_i^a$ is a square matrix like $\begin{bmatrix} G_{i_{11}} & G_{i_{12}} \\ G_{i_{21}} & G_{i_{22}} \end{bmatrix}$. Normally, Kalman filter needs two versions of Kalman gain to update $\hat{x}$ and $\hat{x}^*$. Assuming $\hat{x}$ and $\hat{x}^*$ are of size $m$, this implies that $2m$ states are to be updated. However, in the proposed method, we can form $\hat{x}^*$ based on $\hat{x}$ and deploy $G_{i_{11}}$ and $G_{i_{12}}$ to update the states in (6). This reduces the number of states to be updated to $m$, leading to a substantial decrease in simulation time. In order to decrease the computational time of ACKF, the conjugate parts of updating states expressions can be ignored in equations (5) and (6). Hence, in these two equations augmented state vector is replaced by state vector, while $G_{i_{11}}$ and $G_{i_{12}}$ are deployed to update states based on augmented correction error.

*B. DSE based on ACKF*

In order to develop a DSE-ACKF framework, the injected currents $(i_{inj})$ at different buses are considered as the states $(x_i = i_{inj_i})$ in (1). To consider injected currents as states, transition matrix $F_{i-1}$ should be specified. We conducted a study on the profile of distribution network customer loads and concluded that the difference between the injected currents at two successive time steps can be characterized as a white noise. This allows us to specify $F_{i-1}$ in (2) as an identity matrix. To elaborate this claim, we used measured data of Newmarket suburb, Queensland, Australia with temporal resolution of one minute and for a period of seven days.

In mathematics literature, white noise is defined as a sequence of uncorrelated random variables with zero mean [23]. Fig. 1(a) shows the sequential differences between residential customer injected currents at successive time steps. The red line in this figure represents the mean of the time series, which is almost equal to zero. In addition, the temporal correlation of the difference between one to thirty consecutive time samples of injected current on a rolling base is calculated [24] and visualized in Fig. 1(b). In Fig. 1(b), the black to white colors represents the highest to lowest correlation values. The element in nth row and jth column of the correlation matrix represents the correlation between the injected current differences with n and m minutes lags. Fig. 1(b) implies that injected current differences with different lags are independent. These empirical results allow us to consider $F_{i-1}$ as an identity matrix to ascertain that the difference between two adjacent time samples of injected current is equal to the state noise ($w_i$).

In order to define the measurement vector and the observation matrix, direct power flow algorithm [25] is employed. In this method, the relation between injected currents ($i_{inj}$), branch currents ($i_{branch}$) and bus voltages ($v_{inj}$) are formulated as:

$$\begin{cases} i_{branch} = BIBC * i_{inj} \\ v_{inj} = v_{ref} - DLF * i_{inj} \end{cases} \quad (7)$$

where $BIBC$ is the Bus-Injection to Branch-Current matrix, $DLF$ is the Direct Load Flow matrix and $v_{ref}$ is the voltage of the reference bus [25].

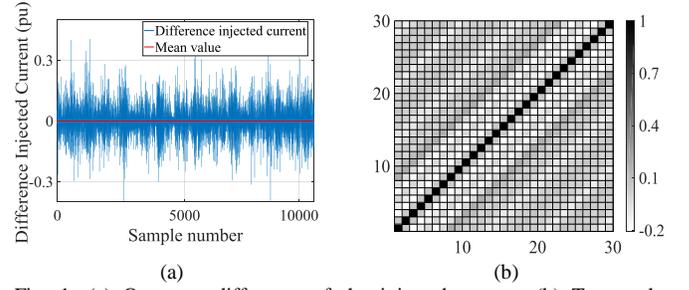

Fig. 1: (a) One step difference of the injected current, (b) Temporal correlation.

The measured vector contains a combination of calculated pseudo data and a few real measurements. The measurement matrix includes, pseudo measured subset of injected currents ($Ip_{inj}^a$), a limited number of measured bus voltages ($Vm_{inj}^a$) and branch currents ($Im_{branch}^a$) as:

$$y_i = \begin{bmatrix} Ip_{inj_i}^a \\ Im_{branch_i}^a \\ Vm_{inj_i}^a \end{bmatrix} \quad (8)$$

Based on (7), the transient matrix $H$ is given as:

$$H = \begin{bmatrix} I_{(m-1)} \\ BIBC_m \\ -DLF_m \end{bmatrix} \quad (9)$$

where $BIBC_m$ and $DLF_m$ contain selected rows of $BIBC$ and $DLF$ matrices for measured branch currents and bus voltages.

The measurement noise covariance matrix represents the dependency between the elements of the real and imaginary parts of the measured vector, and consists of three independent segments as given in (10). $R_{inj_i}^a$ is the covariance matrix of injected currents pseudo measurement noise, The elements of $R_{inj_i}^a$ are high in value due to the low accuracy of pseudo data. Similarly, $R_{branch_i}^a$ and $R_{V_i}^a$ as branch currents and bus voltages covariance matrices whose elements are comparatively lower and they reflect the dependencies in the error of measurement devices.

$$R_i^a = \begin{bmatrix} R_{inj_i}^a & 0 & 0 \\ 0 & R_{branch_i}^a & 0 \\ 0 & 0 & R_{V_i}^a \end{bmatrix} \quad (10)$$

Both snapshot and time series estimators have increasing difficulty in handling large numbers of nodes as the computational effort increases with the cube of bus numbers. Layering can break the problem of the large networks into a set of smaller network problems which solve much faster. In section III, the details of the proposed multi-layer state estimator are provided.

## III. MULTI-LAYER STATE ESTIMATION

The large number of customers in distribution systems makes the DSE techniques computationally extensive. We propose a multi-layer FASE approach to improve computational speed and facilitate distributed solutions. In the proposed framework, each distribution network is divided into one main area and several subareas, as shown in Fig. 2.

In the MV/LV transformer area, voltage and current of the secondary side of the transformer are measured. The states of the main area are injected current of each subarea and the voltages at boundaries between main area and subareas. In the main area, the subareas act like big customer loads in the state estimator. The historical data from temporary measurements are employed as the pseudo measured data of total consumed power in each subarea. However, historical data are not accurate enough for state estimation, yet the injected currents of subareas as the states play an important role in control and protection. Hence, we propose an idea to determine the real-time subareas' injected currents based on the measured current on the secondary of the MV/LV transformer. We calculate the contribution of each subarea to the measured current on the secondary of the MV/LV transformer based on historical data and in real-time compute the injected currents of subareas as:

$$SF_j(t) = \frac{I_j^P(t)}{\sum_{i=1}^n I_i^P(t)} \quad (11)$$

$$I_{inj_j}^{Updated}(t) = SF_j * I_{T_s}^M(t) \quad (12)$$

where $i$ and $j$ are subarea indices, $SF_j$ is the pseudo scaling factor of subarea $j$, $I_{inj_j}^{Updated}$ is the updated injected current of subarea j, $I_j^P$ is the pseudo current values of subarea $j$ from the historical data. $I_{T_s}^M$ is the measured current on the secondary side of transformer and $t$ is the sampling time. It is worth to note that equation (11) is based on the injected currents not consumed power. Hence, the network loss is not neglected and is considered in power flow formulation.

Geographical aggregation has smoothing effects. Therefore, it is expected that the aggregation of loads in each subarea comparing to individual loads shows a lower variability. These make the computed pseudo scaling factor more reliable and lead to a higher accuracy in estimating the updated pseudo data. The measured residential customer loads data taken from Newmarket suburb in Brisbane Australia is employed to compare the variability of the real scaling factor of main branch currents comparing to those of individual customer loads. Fig. 3 visualizes the real scaling factor of the injected current of an individual customer Fig. 3(a) as well as the aggregated injected current of a group of twenty customers Fig. 3(b) at a distribution network with forty customers in one day. As expected, the aggregated loads show much smoother real scaling factor comparing to an individual load. The variations of the real scaling factor of a group of aggregated houses in one day is very low, while the same for an individual customer is high and difficult to predict. Therefore, it is expected that the pseudo scaling factor calculated for a group of aggregated houses using the pseudo data can represent the real scaling factor of the group for future periods while lower estimation accuracy is expected to observe for individual customers.

It is worth noting that aggregation of customers which are in close proximity improves the accuracy of the state estimator. However, the aggregation for those customers that are significantly distant from one another with substantial voltage differences, reduces the accuracy of state estimation. In this paper aggregation does not involve substantial voltage differences.

In order to guide through a straightforward implementation of the proposed DSE-MACKF framework, the flowchart of the proposed algorithm is illustrated in Fig. 4. Also, our proposed approach is explained step-by-step in the following.

- *Step 1:* The required input data is imported. These include pseudo and real measured data, the network schematics and line impedances to calculate *BIBC* and *DLF* matrices.
- *Step 2:* Determine the main area based on their geographical locations and available real measurements, and form the *BIBC* and *DLF* matrices for the main area and subareas.
- *Step 3:* Use the pseudo data in the power flow algorithm (7) to calculate pseudo branch currents.
- *Step 4:* Based on Step 3, find the pseudo scaling factors for the main branches and individual customers using (11).
- *Step 5:* In the main estimation layer, the pseudo-measured data are updated using the results in Step 4, and the main area's states are estimated using DSE-ACKF algorithm.
- *Step 6:* Based on the estimated states in Step 5, the voltages at boundaries between the main area and subareas are calculated using (7).
- *Step 7:* In the sub layers, the results of the estimated branch currents are used to update the customers' injected currents based on the subareas pseudo scaling factors. In (12), $I_{T_s}^M$ is replaced by estimated branch currents.
- *Step 8:* The results of Step 7 and the estimated voltages at boundaries between the main area and subareas from Step 6 are considered for a parallel state estimation. For subareas' parallel state estimation, (13) is considered [26]:

$$[I_{inj}, I_{branch}, V_{inj}]_j = \begin{bmatrix} I_{inj_j}^{Updated} \\ BIBC_j * I_{inj_j}^{Updated} \\ -DLF_j * I_{inj_j}^{Updated} + v_{b_j} \end{bmatrix} \quad (13)$$

with $v_{b_j}$ as boundary voltage of subarea $j$.

The proposed multi-layer one-iteration state estimation algorithm is designed to provide high accuracy tailored for online control and protection applications. In order to decrease the computational time further, the first four steps can be computed off-line, while other steps should be processed in real-time. The off-line and real-time steps are separated, and parallel process for subareas is shown in Fig. 4.

## IV. DESCRIPTION OF DATA

In this study, smart meters used to collect data from one hundred houses with 15%-25% PV penetration rate in Newmarket suburb. Seven days of data with one-minute resolution is used in simulations and the data from the previous day is considered as pseudo data. The differences between injected currents in two consecutive days considered to form the measurement noise covariance matrix based on (3). Based on the available data from temporary measurements or billing data the total consumed energy of each subarea can be calculated to find the pseudo scaling factors. Although these factors improve the accuracy of pseudo data considerably, still the high uncertainty of PV power data should be accounted for. Therefore, $R_{inj_i}^a$ is considered high enough to model PV uncertainties in the proposed algorithm.

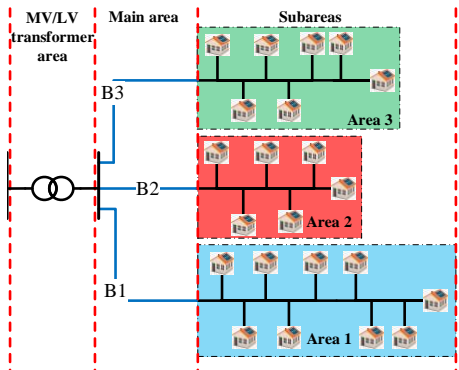

Fig. 2: A typical multi-layer representation of a distribution network.

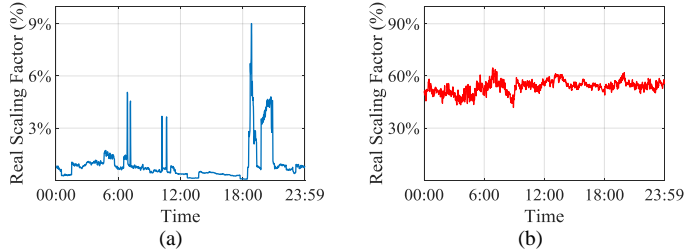

Fig. 3: Contribution percentage (a) individual customer, (b) group of customers.

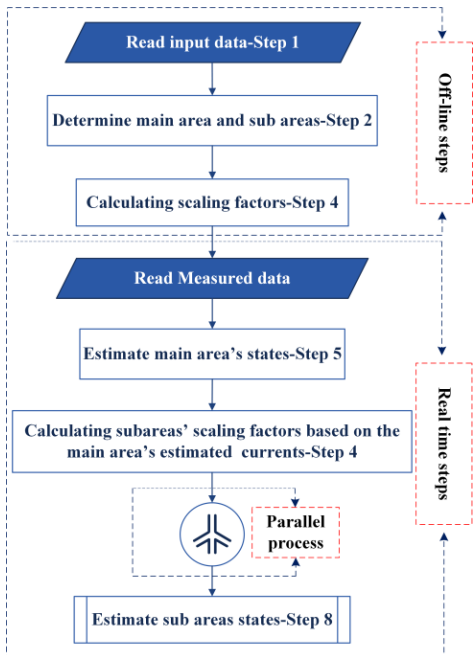

Fig. 4: The flowchart of the layered state estimator.

It is worth to note that this complex FASE requires the measured data with magnitude and angle. However, for distribution network with traditional magnitude measurement devices, the calculated angle based on the pseudo data can be considered for complex state estimation.

In order to detect bad data, the corrective error in (6) is considered as:

$$Error = y_i - H_i \hat{x}_{i|i-1} \quad (14)$$

It is assumed that the corrective error follows a normal distribution with zero mean and a standard deviation ($SD$). For a normally distributed random variable, 99.7% of error values lie within the band of $\pm 3SD$. Usually, the measured data out of this range are considered as bad data. However, the uncertain distribution networks with high PV penetration as well as low a number of measurement devices might distort data from being fully characterized by a normal distribution. Hence, to avoid removing important information from data $\pm 5SD$ corrective error is considered as a reliable marginal band for bad data detection. It is worth noting that, corrective error in bad data detection is monitored for only measured branch currents and bus voltages.

## V. SIMULATION RESULTS

In this section, two case studies are considered. The first case study is a real six bus radial distribution network studied here as a balanced residential network [26]. In the second case study, an unbalanced 23-bus Australian distribution network is considered to evaluate the DSE-MACKF algorithm [27]. The average magnitude voltage error (AMVE), average angle voltage error (AAVE), maximum magnitude voltage error (MMVE), maximum angle voltage error (MAVE), SD of the estimation error in one day, computational time and number of iterations are considered as comparison criteria. AMVE/AAVE and MMVE/MAVE are average and maximum voltage estimation errors in each area, for state estimation simulation over the course of one day. For a comparative evaluation of the proposed algorithm, the accuracy of the estimated states in case study 1 is compared with the results of the WLS algorithm [28].

### A. Case Study 1: 6-bus distribution system

Fig. 5 shows a 6-bus radial distribution network with five residential areas and PV power generation. We use balanced DSE using ACKF for this network. In order to evaluate the effectiveness of the proposed method, a set of one-minute data from the Newmarket suburb is used. In each area, there are twenty houses, and the only available measured parameters are the current and voltage measurements on the LV side of the MV/LV transformer. ACKF and WLS state estimators are employed to estimate all bus voltage magnitudes and angles. As given in Table I, it takes less than 2s for the proposed DSE-based ACKF algorithm to process 1440 samples in one-minute intervals for one-day duration, while WLS needs 14s for the same. The fast performance of the proposed algorithm is a result of its non-iterative nature while WLS requires five iterations in this case. By refining states in the proposed algorithm comparing to WLS the SD of the estimation error improves significantly. For example, in area 4 (with the highest error in estimation) the SD is decreased from 0.0097 to 0.0006. The MMVE computed for this area given by ACKF is 0.22%, while the same is 1.13% using WLS. The AMVE given by the proposed method is near 0.05%, which is one tenth of the average error associated with WLS. Fig. 6 illustrates the efficiency of our method compared to WLS in estimating voltage magnitudes in Case Study 1. The proposed method uses the error at the previous time step as a corrective term to increase the accuracy of the estimated states. Table II summarizes the voltage angle estimation errors in both methods, suggesting that the proposed method delivers a significantly better performance compared with WLS.

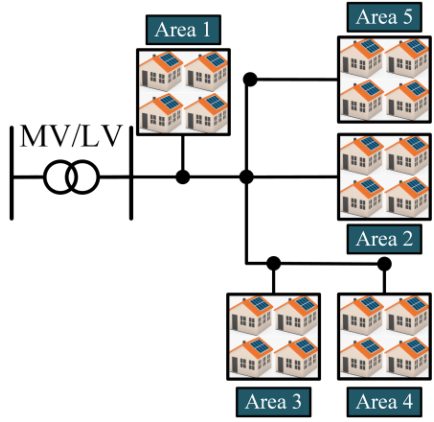

Fig. 5: A six-bus distribution network.

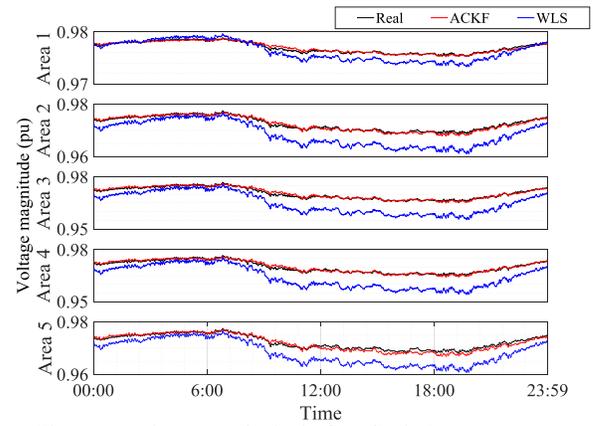

Fig. 6: Five areas voltage magnitudes in Case Study 1.

TABLE I
VOLTAGE MAGNITUDE ERROR IN CASE STUDY 1

|  | WLS | | | ACKF | | |
|---|---|---|---|---|---|---|
|  | t = 14 sec | | | t = 2 sec | | |
|  | AMVE (%) | MMVE (%) | SD (pu) | AMVE (%) | MMVE (%) | SD (pu) |
| Area 1 | 0.09% | 0.20% | 0.0031 | 0.002% | 0.06% | 0.0001 |
| Area 2 | 0.40% | 0.77% | 0.0072 | 0.04% | 0.15% | 0.0003 |
| Area 3 | 0.56% | 1.10% | 0.0093 | 0.04% | 0.22% | 0.0005 |
| Area 4 | 0.60% | 1.13% | 0.0097 | 0.05% | 0.23% | 0.0006 |
| Area 5 | 0.41% | 0.77% | 0.0073 | 0.05% | 0.19% | 0.0004 |

TABLE II
VOLTAGE ANGLE ERROR IN CASE STUDY 1

|  | WLS | | | ACKF | | |
|---|---|---|---|---|---|---|
|  | AAVE (deg) | MAVE (deg) | SD (deg) | AAVE (deg) | MAVE (deg) | SD (deg) |
| Area 1 | 0.227 | 0.350 | 0.066 | 0.005 | 0.017 | 0.005 |
| Area 2 | 0.559 | 0.864 | 0.162 | 0.018 | 0.075 | 0.016 |
| Area 3 | 0.689 | 1.068 | 0.190 | 0.020 | 0.068 | 0.017 |
| Area 4 | 0.720 | 1.12 | 0.200 | 0.023 | 0.085 | 0.017 |
| Area 5 | 0.586 | 0.904 | 0.175 | 0.022 | 0.075 | 0.020 |

In area 4, the MAVE is 0.085 degree in the proposed method, while it is 1.12 degree in WLS. Additionally, the AAVE in the proposed method is around 0.023 degrees, thirty times smaller than AAVE in WLS.

In order to examine the performance of the proposed method, area 1 with the best estimated states, and area 4 with the highest estimation errors are visualized in three-dimensional graphs in Fig. 7. To compare the performances of two estimators in a period with high customer load variations, the real and imaginary parts of the estimated voltages between 5 pm and 8 pm are covered in the figure. The WLS errors in area 1 are lower than those in area 4, due to a shorter feeder length of area 1 from the MV/LV transformer. As the distance increases, the estimation errors in both real and imaginary parts increase. Interestingly, because of complex formulation and regressive process, the proposed method shows a reliable performance with low estimation errors in both areas.

For bad data detection study, a current measurement device is considered on the secondary side of MV/LV transformer. Based on equation (14) the corrective error of historical current measurement data and its corresponding SD are calculated. A $\pm 5SD$ marginal error to detect bad measured data is considered and shown in Fig. 8. According to Fig. 8, no bad data is detected during the one day in the study.

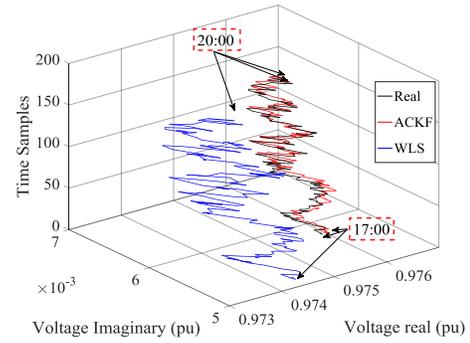

(a)

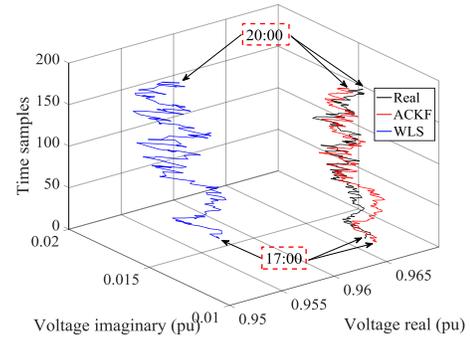

(b)

Fig. 7: The estimated voltages, real and imaginary parts (a) area 1, (b) area 4.

### B. Case Study 2: A 23-bus Australian distribution network

The unbalanced 23-bus distribution network is shown in Fig. 9. The DSE- MACKF algorithm is formed for this network by deploying $BIBC$ and $DLF$ matrices. In order to decrease the simulation time, the network is divided into one main area and three subareas. Subarea 3 has ten buses and imposes a higher computation time compared with other subareas. Hence, by adding another estimation layer, this subarea itself is divided into two subareas, where bus 17 aggregates the loading of subarea 4. Fig. 10 illustrates the proposed algorithm through three layers. In the first layer, the current of main branches and the voltages at boundaries between the main area and subareas are estimated. In the second layer, by incorporating results for bus 2, 10 and 11 from the first layer, the state estimation is conducted in parallel for three independent subareas. Finally, based on the estimated states of bus 17, the states in layer 3 are estimated.

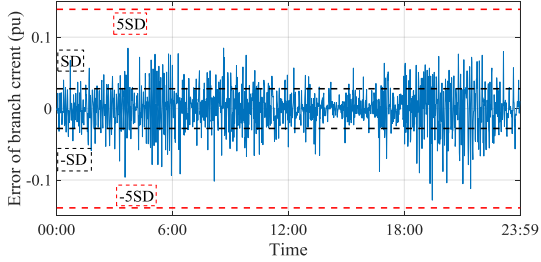

Fig. 8: Corrective error for the current measurement on the secondary side of MV/LV transformer.

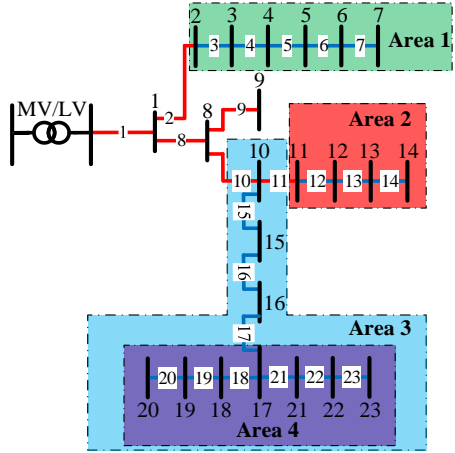

Fig. 9: A MV/LV unbalanced distribution network.

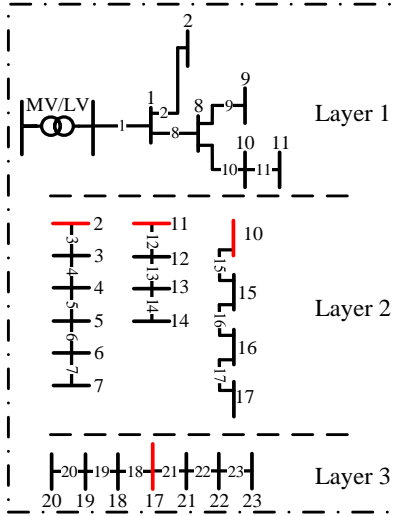

Fig. 10: A three-layer state estimation representation.

For one-day state estimation, the results of one-layer estimation are compared with those of three-layer estimation in Table III. By increasing the accuracy of pseudo scaling factors in the three-layer DSE-MACKF, the MMVE is decreased by 0.13%, while that is 0.57% in the one-layer DSE-MACKF. The major advantage of the three-layer DSE-MACKF is bringing about high computational efficiency. DSE-MACKF reduces the computational time significantly to only 68s, while it is 271s for the one-layer estimation algorithm. The reason behind such improvement is that DSE-MACKF breaks the big matrices to several small ones, and preforms parallel state estimation in the second layer using smaller matrices. Fig. 11 and 12 show the estimated three-phase voltage magnitude for bus 10 and 23. Bus 10 is chosen, because of its important role in estimating the states of subareas 2, 3 and 4. As shown in Table IV, the MVE and AVE for three-layer estimation in bus 10 are 0.32 % and 0.06 %, respectively. These values are low enough to guarantee the accuracy of estimated states in subareas 2, 3 and 4. Additionally, bus 23 is considered as the worst case, due to it is highest distance from the MV/LV transformer. The states of this bus are estimated in the last layer. It is expected that layer three has the highest error among three estimating layers. As shown, in Fig. 12, the magnitude voltage errors in all three phases are fairly acceptable. The MVE is 0.6 and the AVE is 0.12%, while the same metrics are 0.9% and 0.13% respectively, for the one-layer DSE-MACKF.

## VI. CONCLUSION

An accurate and efficient multi-layer DSE-based ACKF algorithm as a FASE method is proposed to estimate nodal voltages based on estimated injected currents. Our investigations showed that the injected currents can be represented as the integral of white noise. This allows us to consider them as the states of the ACKF. Results confirmed that employing DSE-based ACKF even in a poorly monitored distribution network substantially decreases the estimation error. In the proposed DSE-based ACKF, the pseudo scaling factors are modeled to update pseudo-measured data for both individual and groups of residential customer loads. Pseudo scaling factors represent the contribution of each subarea to the measured current on the secondary of the MV/LV transformer. They provide valuable information for FASE to estimate states based on only the voltage and the current measurement devices on the secondary side of the distribution transformer. This significantly reduces the operation cost. In the proposed method, multi estimation layers are considered to perform state estimation hierarchically, leading to considerable fast FASE algorithm for large distribution networks. The idea of phase adjacent aggregation, aggregating loads in subarea leads to observe higher correlations between consumption rates in subareas, when compared to correlation rates between individual loads from two different subareas. This allows us to estimate the scaling factor more accurately because aggregation reduces fluctuations. In addition, the proposed method can deal with complex states and measured values in a reasonable time and result in acceptable error rates. In order to verify the accuracy of the proposed method, two case studies with balanced and unbalanced customer loads, in the presence of PV rooftops are considered. Significant improvements in both simulation time and accuracy of estimation are obtained by deploying multi-layer DSE-based ACKF in both case studies. In Case Study 2, the computational time and complexity of DSE decreased by dividing the distribution network into several subareas and conducting state estimation for subareas in parallel. Incorporation of series and multi-layer parallel process decreased simulation time from 271s in one-layer estimation to 68s in multi-layer case. In the MVE case also decreased from 0.57% to 0.44%.

TABLE III
MAGNITUDE VOLTAGE ERROR IN CASE STUDY 2

|  | One layer ACKF | Three layers ACKF |
|---|---|---|
| AMVE (%) | 0.18 | 0.17 |
| MMVE (%) | 0.57 | 0.44 |
| Time (sec) | 271 | 68 |

TABLE IV
VOLTAGE MAGNITUDE ERROR AND COMPUTATIONAL TIME IN CASE STUDY 2

|  | One layer ACKF | | Three layers ACKF | |
|---|---|---|---|---|
|  | Bus 10 | Bus 23 | Bus 10 | Bus 23 |
| AMVE (%) | 0.06 | 0.13 | 0.06 | 0.12 |
| MMVE (%) | 0.43 | 0.9 | 0.32 | 0.6 |

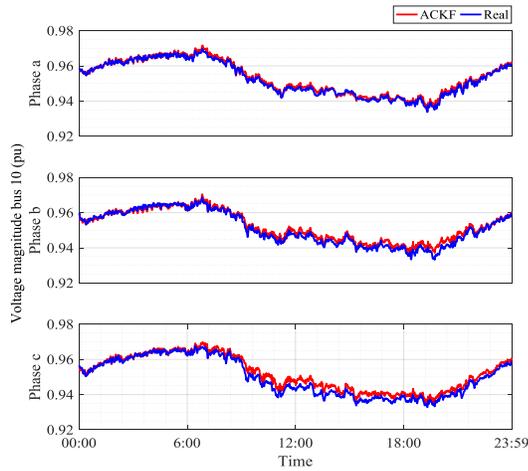

Fig. 11: Three phase voltage magnitudes at bus 10.

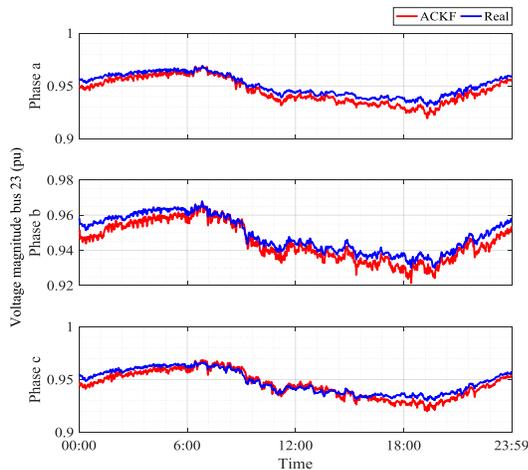

Fig. 12: Three phase voltage magnitudes at bus 23.